\documentclass[onecolumn,journal,twoside]{IEEEtran}

\usepackage[utf8]{inputenc}

\usepackage[T1]{fontenc}
\usepackage{url}
\usepackage{ifthen}
\usepackage{cite}

\usepackage{graphicx}
\usepackage{amssymb}

\usepackage[cmex10]{amsmath} % Use the [cmex10] option to ensure complicance
\usepackage{amsthm}

\newtheorem{theorem}{Theorem}

\newtheorem{lemma}{Lemma}

\theoremstyle{remark}
\newtheorem{remark}{Remark}

\usepackage{xcolor}

 \newcommand{\bF}{\mathbb{F}}
 \newcommand{\bV}{\mathbb{V}}
 \newcommand{\bU}{\mathbb{U}}

\newcommand{\Added}[1]{{#1}}

\begin{document}

\title{%Cascaded? 
Harmonic %?
Coding: An Optimal Linear Code for Privacy-Preserving Gradient-Type Computation}

\author{%
 % \IEEEauthorblockN{Qian Yu}
 % \IEEEauthorblockA{ETH Zürich\\
 %                   ISI (D-ITET), ETH Zentrum\\
 %                   CH-8092 Zürich, Switzerland\\
 %                   Email: moser@isi.ee.ethz.ch}
 % \and
  Qian Yu and A.~Salman~Avestimehr\\
Department of Electrical Engineering, University of Southern California, Los Angeles, CA, USA}

\maketitle

%%%%%%
%% Abstract: 
%% If your paper is eligible for the student paper award, please add
%% the comment "THIS PAPER IS ELIGIBLE FOR THE STUDENT PAPER
%% AWARD." as a first line in the abstract. 
%% For the final version of the accepted paper, please do not forget
%% to remove this comment!
%%
\begin{abstract}
We consider the problem of distributedly computing a general class of functions, referred to as \emph{gradient-type computation}, while maintaining the privacy of the input dataset. Gradient-type computation evaluates the sum of some ``partial gradients'', defined as polynomials of subsets of the input. It underlies many algorithms in machine learning and data analytics.  
We propose Harmonic Coding, which universally computes \emph{any} gradient-type function, while requiring the minimum possible number of workers. Harmonic Coding strictly improves computing schemes developed based on prior works, such as Shamir's secret sharing and Lagrange Coded Computing, by injecting coded redundancy using \emph{harmonic progression}. 
It enables the computing results of the workers to be interpreted as the sum of partial gradients and some redundant results, which then allows the cancellation of non-gradient terms in the decoding process.  
By proving a matching converse, we demonstrate the optimality of Harmonic Coding, even compared to the schemes that are non-universal (i.e., can be designed based on a specific gradient-type function). 
%a class of functions, which can be written as the sum of some . 
%1.  improves
%2. optimal
%3. uses harmonic
%4. arbitrary poly
\end{abstract}

%% The paper must be self-contained. However, if you are referring to
%% a full version for checking certain proofs, please provide the
%% publically accessible location below.  If the paper is completely
%% self-contained, you can remove the following line from your
%% submission.

\section{Introduction}

Gradient computation is the key building block in many optimization and machine learning  algorithms. This computation can be simply described as computing the sum of some ``partial gradients'', which are defined as evaluations of a certain function over disjoint subsets of the input data. This computation structure also broadly appears in various frameworks such as MapReduce \cite{dean2004mapreduce} and tensor algebra \cite{renteln2013manifolds}. We refer to it in general as \textit{gradient-type computation}.
%A commonly used building block in 

%Considering that our targeting application commonly requires

%Computing tasks that  gradient-type computation 
Modern applications that use gradient-type computation often require handling massive amount of data, and distributing the storage and computation onto multiple machines has become a common approach.    
%The massive amount of data and processing power required by 
%Modern computing tasks requires handling increasingly massive amount data, which relies on the computing power and storage provided by distributed computing frameworks. 
However, as more participants come into play, ensuring the privacy of datasets against potential ``curious" workers becomes a fundamental challenge. This critical problem has created a surge of interests in privacy-preserving machine learning (e.g.,~\cite{nikolaenko2013privacy, gascon2017privacy, mohassel2017secureml,chen2019secure}).

\begin{figure}
  \centering
    \includegraphics[width=0.4\textwidth]{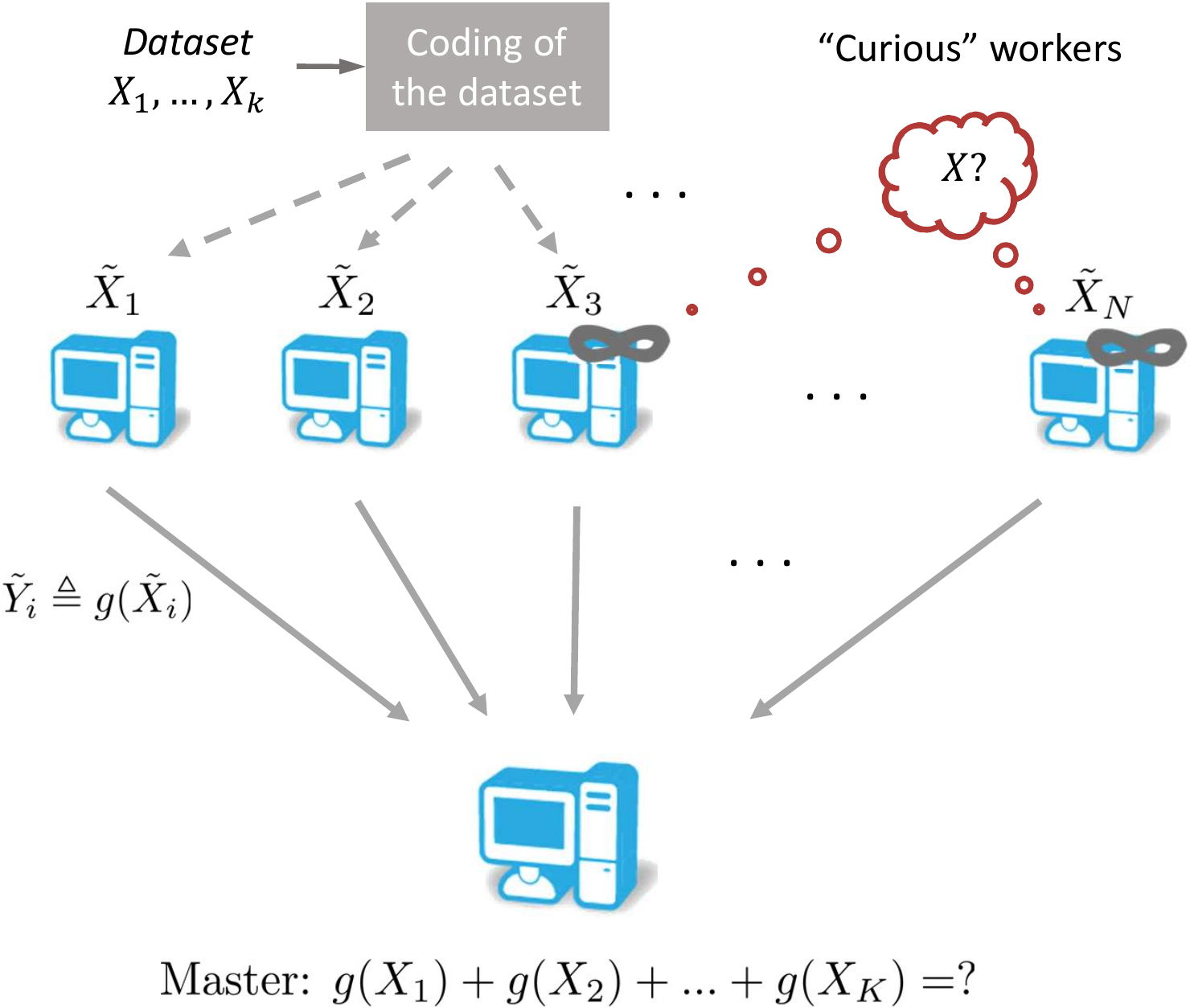}
    \caption{%{\small
    An overview of the framework considered in this paper. 
    The goal is to design a privacy-preserving coded computing scheme for any gradient-type function, using the minimum possible number of workers. The master aims to recover $f(X_1,...,X_K)\triangleq g(X_1)+...+g(X_K)$ given an input dataset $X=(X_1,...,X_K)$, for a \emph{not necessarily linear} function $g$. Each worker $i$ takes a \textit{coded} version of the inputs (denoted by $\tilde{X}_i$) and computes~$g(\tilde{X}_i)$. By carefully designing the coding strategy, the master can decode the function given computing results from the workers, while keeping the dataset private to any \textit{curious worker} (workers~$3$ and $N$ in this example).}
  \label{fig:sys}
\end{figure}

As such motivated, we consider a master-worker computing framework, where the goal is to compute 
\begin{align}\label{eq:def}
    f(X_1,...,X_K)\triangleq g(X_1)+...+g(X_K),
\end{align}
given a large input dataset $X=(X_1,...,X_K)$, where $g$ could be any fixed multivariate polynomial with degree $\textup{deg}\ g$ (See Fig. \ref{fig:sys}). Each worker $i$ can take a coded version of the input variables, denoted $\tilde{X}_i$, and then return $g(\tilde{X}_i)$ to the master. We aim to find an optimum encoding design, which uses the minimum number of worker machines, such that the master can recover $f(X_1,...,X_K)$ given all computing results,  % from all workers, 
and the input dataset is information theoretically private to any worker.

We present a novel coded computing design, called the ``Harmonic coding'', which universally computes any gradient-type function while enabling the privacy of input data. Our main result is that by carefully designing the encoding strategy, the proposed Harmonic Coding computes any gradient-type function while providing the required data privacy using only $N=K(\textup{deg}\ g-1)+2$ workers. This design strictly improves over the state-of-the-art cryptographic approaches that are based on Shamir's secret sharing scheme \cite{Shamir:1979:SS:359168.359176} and the recently proposed Lagrange Coded Computing (LCC) \cite{yu2018lagrange} that can be applied to general polynomial computations. These schemes would respectively  require $N_{\textup{Shamir}}\triangleq K(\textup{deg} \ g+1)$ and $N_{\textup{LCC}}\triangleq K\textup{deg} \ g+1$ workers. Moreover, Harmonic Coding universally computes \emph{any} gradient-type function with any given degree, using \emph{identical} encoding designs. This property allows pre-computing and storing the encoded data far before the identity of the computing task is revealed, reducing the required computation time.

%Harmonic coding uses the minimum possible number of worker machines, and strictly improves 
%over the state-of-the-art cryptographic approaches that are based on Shamir's secret sharing scheme \cite{Shamir:1979:SS:359168.359176} and the recently proposed Lagrange Coded Computing (LCC) \cite{yu2018lagrange} that can be applied to general polynomial computations.    

%Specifically, 

The main idea of Harmonic Coding is by designing the encoding of the dataset,
%each defined as a linearly coded version of the input data, 
 the computation result from each worker can be viewed as a linear combination of a partial gradient and some %evaluations of function $g$ on a list of
 predefined intermediate variables. Using harmonic progression in the coding design, we can let both the intermediate variables and their coefficients be redundant, which enables cancellation of the unneeded results in the decoding process. Moreover, Harmonic Coding has a simple recursive structure,  which allows efficient (linear complexity) algorithms for both encoding and decoding.

 We prove the optimality of Harmonic coding through a matching converse. We show that any linear scheme that computes any gradient-type function with data privacy must use at least the same number of workers required by Harmonic Coding, when the characteristic of the base field is sufficiently large.  In the other case where the characteristic of the base field could be small, we show that improved scheme could be developed for certain functions, while Harmonic Coding remains optimal whenever the partial gradient function $g$ is multilinear. As a side consequence, this converse result also provides a sharp characterization on the characteristic condition of the base field for the existence of universally optimal schemes.

\noindent \textbf{Related Work.} 
Coded computing techniques have seen tremendous success recently in alleviating the communication bottleneck and mitigating stragglers in distributed computing/learning settings (e.g.~\cite{speeding,Li2018fundamental,dutta2016short,yu2017polynomial,codedTerasort,raviv2017gradient,ye2018communication,yu2018straggler,DBLP:journals/corr/abs-1801-10292,li2017nearOptimal,yu2018lagrange}). More recently, coded computing techniques has also been developed for secure and private learning, in particular for logistic regression~\cite{so2019codedprivateml}.

%Conventionally, secret sharing (also lagruange?) and MPC (BGW). LCC, oneshot. modeover, we show that it requires the optimal for the given computing tasks. 

%In this work, we 

%coded computing mostly straggler and fault 

%and various efforts have been made to resolve this challenge.

%Due to the increasing data size of 
%Big data requires distributed computing. Privacy of data is important.  

%data privacy

%1. distributed computing and data privacy? (non can recover)

%1. gradient computation, learning

%3. conventionally: secret sharing, MPC. Coded computing?

\section{Problem Formulation}

We consider a problem of evaluating a gradient-type function $f:\bV^K\to\bU$, characterized by a multivariate polynomial $g:\bV\to\bU$ according to equation (\ref{eq:def}), given an input dataset $X_1,...,X_K\in \bV$, where  $\bV$ and~$\bU$ are some vector spaces over a finite 
field~$\bF$.\footnote{We focus on the non-trivial case, where $g$ is not a constant. We also assume that $\bF$ is sufficiently large.} %with finite dimensions.
The computation is carried out in a distributed system with a master and $N$ workers, where each worker $i$ can take a coded variable, denoted by $\tilde{X}_i$, compute $g(\tilde{X}_i)$, and return the result to the master. The master aims to recover $ f(X_1,...,X_K)$ using all computing results from the workers.

More specifically, using some possibly random \emph{encoding functions} $h_1,...,h_N$, each worker $i$ stores $\tilde{X}_i=h_i(X_1,...,X_K)$ prior to computation. Then after all workers return the results, the master uses a \emph{decoding function} $\ell$ (which is also possibly random, but is independent of the encoding functions) to recover the final output, by computing $\ell(g(\tilde{X}_1),...,g(\tilde{X}_N))$. We restrict our attention to \emph{linear} coding schemes\footnote{A formal definition is provided in Section \ref{sec:converse}. }, which ensures low coding complexities and is easy to implement.

We say a computing scheme is \emph{valid}, if the master always recovers $ f(X_1,...,X_K)$ for any possible values of the input dataset. 
Moreover, we require that the encoding scheme be \emph{data-private}, in the sense that   
%We require that the computing tasks are assigned to 
\emph{none} of the workers can infer any information regarding the input dataset. Formally, we require that 
\begin{align}
    I(X_1,...,X_K;\tilde{X}_i)=0
\end{align}
for any {worker} $i$, if the input variables are randomly sampled from any distribution. 

We aim to characterize the fundamental limit of this problem: finding the minimum possible number of workers among all valid data-private encoding-decoding designs, as well as finding an explicit construction which achieves this optimality.

%To make sure that the data privacy requirement is well defined, we assume the base field and the input vector space are finite. Moreover, we focus on the case where the base field is sufficiently large.

%linear encoding and linear 

\section{Main Results}

We summarize our main results in the following theorem. Our main result is two fold: we first characterized the number workers required by the proposed Harmonic coding scheme, then we prove that it achieves the fundamental limit, i.e., using the minimum number of workers. 

\begin{theorem}%[Achievability]
\label{thm:ach}
For any gradient-type function characterized by a polynomial $g$,  Harmonic coding provides a data-private scheme using $K(\textup{deg}\ g-1)+2$ workers, where $\textup{deg} \ g$ denotes the total degree\footnote{ Formally, $\textup{deg} \ g$ is defined based on the canonical representation of $g$, in which the individual degree within each term is no more than $(|\mathbb{F}|-1)$.
} of $g$.
%\end{theorem} 
%\begin{theorem}[Converse]
Moreover, Harmonic Coding requires the optimum number of workers among all linear coding schemes, when the characteristic of the base field is greater than $\textup{deg}\ g$. %I.e.,
%Optimal if chara is large?
\end{theorem}

\begin{remark}
To prove the first result in Theorem \ref{thm:ach}, we present Harmonic Coding in Section \ref{sec:achieva}, which only uses the stated optimum number of workers. Compared to Harmonic Coding, one conventional approach used in Multiparty computing (MPC) is to first encode each input variable separately using Shamir's secret sharing scheme\cite{Shamir:1979:SS:359168.359176}, then apply the computation on top of the shared pieces. This MPC based approach requires $\textup{deg} \ g+1$ evaluations of function $g$ to compute every single $g(X_i)$, and thus uses $K(\textup{deg} \ g+1)$ workers in total. More recently, we proposed Lagrange Coded Computing \cite{yu2018lagrange}, which enables evaluating all $g(X_i)$'s through a joint computing design. Lagrange Coded Computing encodes the data by constructing a degree $K$ polynomial whose evaluations are the input variables and the padded random keys at $K+1$ points, then assign the workers its evaluation at $N$ other distinct points. After computation, the decoding process reduces to interpolating a polynomial of degree $K\textup{deg} \ g$, which requires $K\textup{deg} \ g+1$ workers. 
\Added{Harmonic Coding strictly improves these two approaches. }

\Added{Moreover, Harmonic Coding also enables significant reduction in terms of the required number of workers.  As an example, consider the linear regression model presented in \cite{yu2018lagrange}, where the computational bottleneck is to evaluate a gradient-type function characterized by $g(X_i)=X_i^\intercal X_i w$ for some fixed matrix $w$. As the size of the dataset ($K$) increases, the MPC and LCC based designs require approximately $3K$ and $2K$ workers respectively. On the other hand, Harmonic Coding only requires about $K$ workers, which is a two-fold improvement upon the state of the art. Furthermore, it even approaches the fundamental limit for simply storing the data, where no computation is required.      }

\Added{Unlike prior works, our main coding idea is instead by carefully designing the encoding, that the workers compute the sum of the $g(X_i)$'s ``in the air''.} Specifically, we interpret the workers computing results as the sum of a ``partial gradient'' (i.e., a single $g(X_i)$) and some intermediate variables. By encoding the inputs using harmonic progression, all intermediate variables cancels out in the decoding process, and the master directly obtain the sum of all gradients.
\end{remark} 

\begin{remark}
Harmonic coding also has several additional properties of interest. First, Harmonic coding is \emph{identical} for any function $f$ with a given degree. Hence, it enables pre-encoding the data without knowing the identity of the function, and \emph{universally} computes any function with an upper bounded degree. Second, Harmonic Coding enables linear complexity algorithms for both encoding and decoding, hence requiring negligible computational coding overheads for most applications. Finally, to provide data privacy against every single worker, Harmonic Coding only uses one single random key through the entire process. This achieves the minimum amount of randomness required by any linear scheme. 
\end{remark} 

%\begin{remark}
%decoding? randomness
%\end{remark} 

\begin{remark}
Harmonic coding reduces to Lagrange Coded Computing in several basic cases. For example, when $K=1$, the master only needs a single evaluation of function $g$, which can be optimally computed by LCC. On the other hand, when the computation task is linear, due to commutativity between $g$ and the sum in function $f$, the master essentially wants to recover $g(\frac{X_1+...+X_K}{K})$, which can also be optimally computed using LCC by pre-encoding the dataset into a single variable $\frac{X_1+...+X_K}{K}$. However, for all other cases, Harmonic Coding achieves the optimum number of workers, which was earlier unknown.  
\end{remark} 

We complete the proof of Theorem \ref{thm:ach} by providing a matching converse, which is presented in Section \ref{sec:converse}.
%The second converse, idea?

%discussion: the first non-trivial case

%assume linear encoding and decoding

%$T=0$, $d=1$, $N*=1$

%$T=0$, $d>1$, $N*=K$ for multilinear, 

%or general function for any large q?

%dT+(d-1)K

%now $T>0$

%$d=1$, $N^*=T+1$

%$d>1$, for multilinear

\section{Achievability Scheme} \label{sec:achieva}
In this section, we prove the achievability part of Theorem \ref{thm:ach} by presenting Harmonic Coding. We start with a motivating example for the first non-trivial case, where the master aims to recover the sum of $K=2$ quadratic functions.
%We start by presenting Harmonic coding through an motivating example.

\subsection{Example for $K=2$, $\textup{deg} \ g=2$}\label{sec:example}

Consider a gradient-type function given input variables $X_1,X_2\in\mathbb{F}_{5}^{m\times m}$ for some integer $m$, characterized by a quadratic polynomial $g(X_i)=AX_i^{\intercal}X_i+BX_i+C$ with some constant matrices $A$, $B$, and $C$. We aim to find a data-private computing scheme which only uses $4$ workers.

To achieve the privacy requirement, we pick a uniformly random matrix $Z\in\mathbb{F}_{5}^{m\times m}$, and assign the coded variables by linearly combining $X_1$, $X_2$, and $Z$. One can verify that this requirement is satisfied as long as variable $Z$ is encoded with non-zero coefficients for every $\tilde{X}_i$. Hence, it remains to design the code with validity. 

The main idea of Harmonic Coding is to carefully design the linear combinations, such that after applying $g$ to the coded variables, each computing result equals to the sum of a ``partial gradient'' and some intermediate values that can be canceled out in the decoding process. Such property is achieved by encoding the variables using harmonic progression\Added{\footnote{Explicitly, using the sequence $\{\frac{1}{c-i}\}_{i\in\mathbb{N_+}}$ as encoding coefficients.}}. 

Specifically, we first define some parameters, letting $c=4$ and $\beta=4$. These values are selected such that $c\not\in\{0,1,2\}$ and $\beta\not\in\{0,1,\frac{c}{c-1},\frac{c}{c-2}\}$. We then defined some intermediate variables, which are coded using harmonic progression.
\begin{align}
    P_0&\triangleq\frac{c}{c-0}Z=Z \nonumber\\
    P_1&\triangleq\frac{c}{c-1}Z-\frac{1}{c-1}X_1=3X_1+3Z \nonumber\\
    P_2&\triangleq\frac{c}{c-2}Z-\frac{1}{c-2}(X_1+X_2)=2X_2+2X_1+2Z.\nonumber
\end{align}
Given these definitions, the input data is encoded as follows.
%We encode the input dataset and a padded uniformly random variable $Z\in\mathbb{F}_{5}^{m\times n}$ as follows.
\begin{align}
    \tilde{X}_1&=P_0=Z\nonumber\\
    \tilde{X}_2&=(1-\beta)X_1+\beta P_0=2X_1+4Z\nonumber\\
    \tilde{X}_3&=(1-\frac{c-1}{c}\beta)X_2+\frac{c-1}{c}\beta P_1=3X_2+4X_1+4Z\nonumber\\
    \tilde{X}_4&=P_2=2X_2+2X_1+2Z.\nonumber
\end{align}

%\begin{align}
%    \tilde{X}_1&=3X_2+4X_1+4Z\\
%    \tilde{X}_2&=2X_2+2X_1+2Z\\
%    \tilde{X}_3&=2X_1+4Z\\
%    \tilde{X}_4&=Z.
%\end{align}
%As one can see, any of the coded variable is masked the padded variable $Z$, which guarantees the data privacy.

Using this encoding design, the master can decode the final result by computing $2g(\tilde{X}_1)+g(\tilde{X}_2)+3g(\tilde{X}_3)+g(\tilde{X}_4)$, which exactly recovers $g(X_1)+g(X_2)$. As mentioned earlier, the intuition for this decodability is that $g(\tilde{X}_2)$ and $g(\tilde{X}_3)$ can be represented as the sum of some intermediate values and $g(X_1)$, $g(X_2)$, respectively. This representation is constructed using Lagrange's interpolation formula by viewing each of them as a quadratic function of $\beta$ or $\frac{c-1}{c}\beta$.

For instance, by viewing $\tilde{X}_2$ as a linear function of $\beta$, after applying polynomial $g$, we obtain a quadratic function. Reevaluating this function at point $0$ and $1$ gives $g(X_1)$ and $g(P_0)$. Moreover, Harmonic Coding provides a recursive relation $P_1=(1-\frac{c}{c-1})X_1+\frac{c}{c-1}P_0$, which indicates that $g(P_1)$ can be viewed as evaluating the same polynomial at point $\frac{c}{c-1}$. Hence, the Lagrange's interpolation formula suggests
\begin{align}\label{eq:ex1}
    g(X_1)=&\frac{1\cdot \frac{c}{c-1}}{(1-\beta)(\frac{c}{c-1}-\beta)}g(\tilde{X}_2) \nonumber \\ &+ \frac{\beta\cdot \frac{c}{c-1}}{(\beta-1)(\frac{c}{c-1}-1)}g(P_0)\nonumber \\&+ \frac{1\cdot \beta}{(1-\frac{c}{c-1})(\beta-\frac{c}{c-1})}g(P_1) \nonumber \\
    =&g(\tilde{X}_2)+2g(P_0)+3g(P_1).
\end{align}
Similarly, by viewing $g(\tilde{X}_3)$ as a quadratic function of $\frac{c-1}{c}\beta$, and viewing $g(X_2)$, $g(P_1)$ and $g(P_2)$ as its evaluations at points $0$, $1$, and $\frac{c-1}{c-2}$, we have  
\begin{align}\label{eq:ex2}
    g(X_2)=&\frac{1\cdot \frac{c-1}{c-2}}{(1-\frac{c-1}{c}\beta)(\frac{c-1}{c-2}-\frac{c-1}{c}\beta)}g(\tilde{X}_3) \nonumber \\ &+ \frac{\frac{c-1}{c}\beta\cdot \frac{c-1}{c-2}}{(\frac{c-1}{c}\beta-1)(\frac{c-1}{c-2}-1)}g(P_1)\nonumber \\&+ \frac{1\cdot \frac{c-1}{c}\beta}{(1-\frac{c-1}{c-2})(\frac{c-1}{c}\beta-\frac{c-1}{c-2})}g(P_2) \nonumber \\
    =&3g(\tilde{X}_3)+2g(P_1)+g(P_2).
\end{align}

Note that the LHS of equations (\ref{eq:ex1}) and (\ref{eq:ex2}) are exactly the two needed ``partial gradients'', while the sum of the coefficients of $g(P_1)$ on the RHS is zero (which is also due to the Harmonic Coding structure). Thus, by adding these two equations, the intermediate value $g(P_1)$ is canceled, and we have shown that $f(X_1,X_2)$ can be recovered from   $g(\tilde{X}_2)$, $g(\tilde{X}_3)$, $g(P_0)$, and $g(P_2)$. Moreover, note that $g(P_0)$ and $g(P_2)$ are directly computed by worker $1$ and worker $4$. This completes the intuition for the validity of the proposed design.
%This explains the fact that  $f(X_1,X_2)$ can be decoded only using $g(\tilde{X}_1),...,g(\tilde{X}_4)$. 

%After each worker applies function $g$ on the stored coded variable and returns the result to the master, the final output can be decoded by computing , which directly gives $f(X_1,X_2)$.

%aim to find a computing design that preserves data privacy
%From LCC, for any $\alpha,\beta\neq 0,1$, we can encode any two variables $X$, $Y$, such that 
%\begin{align}
%   \frac{1(1-\beta)}{\alpha(\alpha-\beta)}g(\tilde{X}_\alpha)+\frac{1(1-\alpha)}{\beta(\beta-\alpha)}g(\tilde{X}_{\beta})=f(X)-\frac{(1-\alpha)(1-\beta)}{(-\alpha)(-\beta)}f(Y).
%\end{align}
%We can find new variables $\alpha'$, $\beta'$ such that $-\frac{(1-\alpha)(1-\beta)}{(-\alpha)(-\beta)}=\frac{1(1-\beta')}{\alpha'(\alpha'-\beta')}$, then by substituting $Y$ by a coded version of two new random variables $X'$ and $Y'$, we can obtain
%\begin{align}
%  f(X)+f(X')+cf(Y'),
%\end{align}
%for some constant $c$ by linearly combining $3$ computing results. Repeating the same step we obtain a scheme that uses $K+2$ machines.

%scales linearly to the length of each input vector $X_i$ 

\subsection{General Scheme}

Now we present Harmonic Coding for any gradient-type function $f$  with degree $d$, and for any parameter value of $K$. We first partition the workers into $K+2$ groups, where the first $K$ groups each contain $d-1$ workers, and the rest two groups each contains $1$ worker. For brevity, we refer to the workers in the first $K$ groups as ``worker $i$ in group $j$'' for $i\in[d-1]$, $j\in[K]$, and denote the assigned coded variable $\tilde{X}_{(i,j)}$; we refer to the rest $2$ workers as worker $1$ and worker $N$.

Recall that the base field is assumed to be sufficiently large, we can find a parameter $c\in\bF$ that is not from $\{0,1,...,K\}$. Moreover, we find parameters $\beta_1,...,\beta_{d-1}\in\bF$ with distinct values that are not from $ \{0\}\cup \{\frac{c}{c-i}\}_{i=0}^{K}$.  Similar to Section \ref{sec:example}, we use a uniformly random variable $Z\in\bV$, and define intermediate variables $P_0,...,P_K$ as follows.
\begin{align}
P_j\triangleq\frac{c}{c-j}Z-\frac{1}{c-j}\sum_{k=1}^{j} X_K.
\end{align}
Then the input data is encoded based on the following equations. 
\begin{align}
 \tilde{X}_{1}&=P_0\\
    \tilde{X}_{(i,j)}&=X_j(1-\beta_i\frac{c-j+1}{c})+P_{j-1}\beta_i\frac{c-j+1}{c}\label{eq:coded}\\
    \tilde{X}_{N}&=P_K,
\end{align}
%where $\{P_j\}_{j=0}^{K}$ are some intermediate variables defined as follows. %\footnote{For convenience, we  define $P_0=Z$.}

Using the above encoding scheme, one can verify that all coded variables are masked by the variable $Z$, which guarantees the data privacy. Hence, it remains to prove the decodability, i.e., $f(X_1,...,X_K)$ can be recovered by linearly combining the results from the workers.
The proof relies on the following lemma, which is proved in Appendix A.
\begin{lemma}\label{lemma:ach}
For any gradient-type function with degree of at most $d$, using Harmonic coding, the master can compute 
\begin{align}\label{eq:lemma1}
  Q_j\triangleq & g(X_j)- g(P_{j-1}) (c-j+1)\prod_{i=1}^{d-1}\frac{\beta_i(c-j+1)}{\beta_i(c-j+1)-c}\nonumber\\&+g(P_{j})(c-j)\prod_{i=1}^{d-1}\frac{\beta_i(c-j)}{\beta_i(c-j)-c}
\end{align}
for any $j\in[K]$, by linearly combining computing results from workers in group $j$.
\end{lemma}

Similar to the motivating example, the proof idea of Lemma \ref{lemma:ach} is to view $\{g(X_{(i,j)})\}_{i\in[d-1]}$, $g(X_j)$, $g(P_{j-1})$, and $g(P_{j})$ as evaluations of a degree $d$ polynomial at $d+2$ different points, and derive equation (\ref{eq:lemma1}) using Lagrange's interpolation formula. 
Assuming the correctness of the Lemma \ref{lemma:ach}, the master can first decode $Q_j$'s given the computing results from the workers. Then note that in equation (\ref{eq:lemma1}) the sum of the coefficients of each  $g(P_{j})$ in $Q_j$ and $Q_{j+1}$ is zero for any $j\in[K-1]$. By adding all the variables $Q_1,...,Q_K$, the master can obtain a linear combination of $f(X_1,...,X_K)$, $g(P_0)$, and $g(P_K)$. Finally, because $g(P_0)$ and $g(P_K)$ are computed by worker $1$ and worker $N$, $f(X_1,...,X_K)$ can be computed by subtracting the corresponding terms. This proves the validity of Harmonic Coding. 
%\begin{align}
%    \sum_{i=1}^K Q_i&= \sum_{i=1}^K g(X_i)-g(Z)c\prod_{i=1}^{d-1}\frac{\beta_i}{1-\beta_i}+g(P_K)\prod_{i=1}^{d-1}\frac{\beta_i(c-K)}{c-\beta_i(c-K)}\\
%    &=f(X_1,...,X_K)-g(\tilde{X}_1)c\prod_{i=1}^{d-1}\frac{\beta_i}{1-\beta_i}+g(\tilde{X}_N)\prod_{i=1}^{d-1}\frac{\beta_i(c-K)}{c-\beta_i(c-K)}.
%\end{align}
%The master can hence recover $f(X_1,...,X_K)$ by computing the sum of $Q_i$'s, the subtracting the corresponding terms based on computing results from worker $1$ and worker $N$.

%\begin{align}
 %   g(X_i)=g(P_i)\prod_{j\neq i} \frac{\beta_j}{1-\beta_j}+\sum_{k=1}^{d}g(X_i(1-\beta_i)+P_i\beta_i)\left(\prod_{j\neq i} \frac{\beta_j}{\beta_i-\beta_j}\right)\frac{1}{\beta_i-1}
%\end{align}

%idea use less than LCC

%\begin{remark}
%any field size at least $K+d-1$
%\end{remark}

\begin{remark}
Harmonic Coding enables efficient encoding and decoding algorithms, both with linearly complexities. A linear complexity encoding algorithm can be designed by exploiting the recursive structures between the intermediate variables and coded variables. Specifically, the encoder can first compute all $P_i$'s recursively, each $P_i$ by linearly combining $P_{i-1}$ and $X_i$; then every other coded variable that is not available can be computed directly using equation (\ref{eq:coded}). This encoding algorithm requires computing $O(N)$ linear combinations of two variables in $\bV$, which is linear with respect to the output size of the encoder. On the other hand, the decoding process is simply linearly combining the outputs from all workers, and a natural algorithm achieves linear complexity with respect to the input size of the decoder.        
\end{remark}

\begin{remark}
%Harmonic Coding can also be extend to scenarios where the base field $\bF$ is infinite. Note that any practical (digital) implementation for such computing tasks requires quantizing the variables into discrete values. We can thus embed them into a finite field that covers the range of possible final outputs, then directly apply Harmonic Coding and compute the finite field version of the gradient-type function.  This approach also avoids potentially large intermediate computing results, which could save storage and computation time, and reduce numerical errors.
Harmonic Coding can also be extend to scenarios where the base field $\bF$ is infinite. Note that any practical (digital) implementation for such computing tasks requires quantizing the variables into discrete values. We can thus embed them into a finite field, then directly apply the finite field version of Harmonic Coding. \Added{For instance, if the input variables and the coefficients of $g$ are quantized into $k$-bit integers, the length of output values are thus bounded by $\tilde{O}(k \cdot\textup{deg} g)$, which only scales logarithmically with respect to parameters such as number of  workers ($N$). We can always find a finite field $\mathbb{F}_p$ with a prime $p=\tilde{\Theta}(k \cdot\textup{deg} g)$, that enables computing $f$ with zero numerical error. }
%As long as the field covers the range of possible final outputs, Harmonic Coding would return the correct value.
This approach also avoids potentially large intermediate computing results, which could save storage and computation time. 
\end{remark}

\section{Converse}\label{sec:converse}
In this Section, we prove the converse part of Theorem \ref{thm:ach}, which shows the optimality of Harmonic Coding. Formally, we define that \emph{linear coding schemes}
are ones that uses \emph{linear encoding functions} and \emph{linear decoding functions}. A linear encoding function computes a linear combination of the input variables and a list of independent uniformly random keys; while a \emph{linear decoding function} computes a linear combination of workers' output. 

We need to prove that for any gradient-type function characterized by a polynomial $g$, any linear coding scheme requires at least $K(\textup{deg}\ g-1)+2$ workers, if the characteristic of $\bF$ is greater than $\textup{deg}\ g$.
 The proof rely on the following key lemma, which essentially states the optimallity of Harmonic Coding among  any scheme that uses linear encoding functions when $g$ is  multilinear.   
 \begin{lemma}\label{lemma:mul}
 For any gradient-type computation where $g$ is a multilinear function, any valid data-private scheme that uses linear encoding functions requires at least $K(\textup{deg}\ g-1)+2$ workers.
  \end{lemma}
   %The proof of Lemma \ref{lemma:mul} can be found in Appendix \ref{app:pl_lemma2}, and the main idea is to construct instances of input values for any assumed scheme that uses a smaller number of workers, and show that such scheme fails to satisfy validity in these instances.
   The proof of Lemma \ref{lemma:mul} can be found in Appendix \ref{app:pl_lemma2}, and the main idea is to construct instances of input values for any assumed scheme that uses a smaller number of workers, where the validity does not hold true.
   Assuming the correctness of Lemma \ref{lemma:mul} and to prove the converse part of Theorem \ref{thm:ach}, we need to generalize this converse to arbitrary polynomial functions $g$, using the extra assumptions on linear decoding and large characteristic of $\bF$. 
   
   Note that the minimum number of workers stated in Theorem \ref{thm:ach} only depends on the degree of the computation task. We can generalize Lemma \ref{lemma:mul} by showing that for any gradient-type function $f$ that can be computed with $N_f$ workers, there exists a gradient-type function $f'$ with the same degree characterized by a multilinear function, which can also be computed with $N_f$ workers. %for which a valid and data-private linear coding scheme can be found also using $N_f$ workers.
%. For clarity, in this section (and relative appendices) we let $K^*_{f}(K,N)$ denote the minimum recovery threshold given $f$, $K$, and $N$.
%ower bounding the minimum recovery threshold 
%using the  can be used for computing a  for any  leads to 
%first prove the converse  
%To demonstrate the main proof ideas, we start by proving this converse for the special case of multi-linear functions. We summarize this partial result in the following lemma:

Specifically, given any function $f$ characterized by a polynomial $g$ with degree $d$, we provide an explicit construction for such $f'$, which is characterized by a multilinear map  $g':\bV^d\rightarrow \bU$, defined as  %, of which the ``partial gradient function'' denoted by $g'$ is a multilinear function.
%which achieves the same requirements. 
%The construction needs to satisfy certain properties. 
%We state the construction and its property
%stated in the following lemma (proved in Appendix \ref{app:pl_basic}):
 %a non-zero, multilinear function $f'$ with the same degree such that satisfies the above requirement. More specifically, given any computation scheme that tolerates $S$ stragglers for $f$, we show that  a computation design can be constructed for $f'$ that 
 
 %, and we aim to prove $K^*_{f}(K,N)\geq K^*_{f'}(K,N)$ by constructing a computation design of $f'$ that is based on a computation design of $f$ and achieves the same recovery threshold. The construction and its properties are stated in the following lemma (which is proved in Appendix \ref{pl:basic}):
 \begin{align}
    g'(X_{i,1},...,X_{i,d})=\sum_{\mathcal{S}\subseteq[d]}{(-1)^{|\mathcal{S}|}g(\sum_{j\in\mathcal{S}}X_{i,j})}
 \end{align}
  for any $\{X_{i,j}\}_{j\in[d]}\in\bV^{d}$. As we have proved in an earlier work (see Lemma 4 in \cite{yu2018lagrange}),
 $g'$ is multilinear with respect to the $d$ inputs. Moreover, if the characteristic of $\mathbb{F}$ is greater than $d$, then $g'$ is non-zero.

Given this construction, it suffices to prove that $f'$ enables computation designs that uses at most the same number of workers compared to that of $f$. 
We prove this fact by constructing such computing schemes for $f'$ given any design for $f$, presented as follows.

Let $X'_1,...,X'_K\in \bV^{d}$ denote the input variables of $f'$, we use a uniformly random key, denoted $Z'\in \bV^{d}$, and encode these variables linearly using the same encoding matrix used in the scheme for $f$. Then similarly in the decoding process, we let the master compute the final result using the same coefficients for the linear combination.   

Because the same encoding matrix is used, the new scheme  constructed for $f'$ is also data private. On the other hand, note that $g'$ is defined as a linear combination of functions $g(\sum_{j\in\mathcal{S}}X_{i,j})$, each of which is a composition of a linear map and $g$. Given the linearity of the encoding design, for any subset $\mathcal{S}$, the variables   $\{\sum_{j\in\mathcal{S}}X_{i,j}\}_{i\in[K]}$ are encoded as if using the scheme for $f$. Hence, the master would return $\sum_{i\in[K]}g(\sum_{j\in\mathcal{S}}X_{i,j})$ if the workers only evaluate the term corresponding to $\mathcal{S}$. 
%any computation scheme of $g$ can be directly applied to any of these functions. 
Now recall that the decoding function is also assumed to be linear. The same scheme is also valid to any linear combinations of them, which includes $f'$. 
Hence, the same number of worker $N_f$ achievable for $f$ can also be achieved by $f'$. This concludes the converse proof. %Theorem \ref{lemma:resiliency} follows immediately.

\section{Conclusion}
In this paper, we characterized the fundamental limit for computing gradient-type functions distributedly with data privacy. We proposed Harmonic Coding, which uses the optimum number of workers, and proved a matching converse. However, note that by relaxing the assumptions we made in the system model and the converse theorem (e.g., random key access, coding complexity, characteristic of $\bF$), improved schemes could be found. We present the following two examples.\footnote{Similar examples and discussions can also be made to the polynomial evaluation problem we formulated in \cite{yu2018lagrange}.}

%we blah?

\subsection{Random Key Access and Extra Computing Power at Master}

Recall that in the system model, we assumed that the decoding function is independent of the encoding functions. This essentially states that the master does not have access to the random keys when decoding the final results. Moreover, we assumed linear decoding, which restricts the computational power of the master.

However, if both of these assumptions are removed, and the master has the knowledge of function $f$, an improved yet practical scheme based on Harmonic Coding can be obtained by letting the master compute  $g(Z)$ in parallel  with the workers. In this way, the required number of workers can be reduced by $1$. Alternatively, if the master has access to the input data, it can compute any $g(X_i)$, effectively reducing $K$ by $1$, and reducing the required number of workers by $\textup{deg} \ g-1$. %todo check

\subsection{An Optimum Scheme for  $\textup{Char}\ \bF= \textup{deg}\ g$}

The converse presented in Theorem \ref{thm:ach} is only stated for the case where the characteristic of the base field $\bF$ is greater than the degree of function $g$. In fact, when this condition does not hold, improved coding designs can be found for certain functions $f$.
For example, consider a gradient-type function characterized by $g:\bF^m\to\bF^n$ defined as
\begin{align}
    g(X_i)= A \cdot \begin{bmatrix}
    X_{i,1}^d  \\
    X_{i,2}^d  \\
    \vdots \\
    X_{i,m}^d 
\end{bmatrix},
\end{align}
where $A$ is a fixed non-zero $n$-by-$m$ matrix, and $d$ equals the characteristic of $\bF$. By exploiting the ``Freshman's dream'' formula (i.e., $\sum_i X_{i,j}^d=(\sum_i X_{i,j})^d$), one can actually design an optimal data private scheme that only uses $2$ workers for any possible $d$, instead of using Harmonic coding which requires $K(d-1)+2$ workers.   

Recall that Lemma \ref{lemma:mul} states that Harmonic Coding does achieve optimality for any multilinear $g$ and for any characteristic of $\bF$. This in fact shows that when the characteristic of the base field equals $\textup{deg}\ g$, we are not able to find a coding scheme that is universally optimal for \emph{all} gradient-type functions of the same degree. Equivalently, we have shown that the $\textup{Char}\ \bF> \textup{deg}\ g$ requirement in the converse statement in Theorem \ref{thm:ach} provides a sharp lower bound on the characteristic of the base field to guarantee the existence of universally optimal schemes.

%sharp bound

%interestingly even when the field is infinitely large

	\section*{Acknowledgement}
	This material is based upon work supported by Defense Advanced Research Projects Agency (DARPA) under Contract No. HR001117C0053, ARO award W911NF1810400, NSF grants CCF-1703575, ONR Award No. N00014-16-1-2189, and CCF-1763673. The views, opinions, and/or findings expressed are those of the author(s) and should not be interpreted as representing the official views or policies of the Department of Defense or the U.S. Government.  Qian Yu is supported by the Google PhD Fellowship.

	\bibliographystyle{ieeetr}
\bibliography{references}

\appendices

\section{Proof of Lemma \ref{lemma:ach}}\label{app:pl_lemma1}

To prove Lemma \ref{lemma:ach}, it suffices to find a linear combination of $\{g(\tilde{X}_{(i,j)})\}_{i\in[d-1]}$ for each $j\in[K]$ that computes $Q_j$.
As mentioned in Section \ref{sec:achieva}, we construct this function by viewing  $\{g(X_{(i,j)})\}_{i\in[d-1]}$, $g(X_j)$, $g(P_{j-1})$, and $g(P_{j})$ as evaluations of a degree $d$ polynomial at $d+2$ distinct points.

Specifically, we view the coded variables $\{X_{(i,j)}\}_{i\in[d-1]}$ as evaluations of a scalar input linear function, each at point  $\beta_i\frac{c-j+1}{c}$, which evaluates $X_j$ and $P_{j-1}$ at $0$ and $1$. One can verify that the same linear function gives $P_{j}$ at point $\frac{c-j+1}{c-j}$. Moreover, these $d+2$ evaluations are at distinct points due to the requirements we imposed on the values of $c$ and $\beta_i$'s.

After applying function $g$, the corresponding results become evaluations of a degree $d$ polynomial at $d+2$ points. Hence, any one of them can be interpolated using the other $d+1$ values. By interpolating $g(X_j)$, we have   
\begin{align}\label{eq:ex12}
    g(X_j)=&\sum_{i\in[d-1]}g(\tilde{X}_{(i,j)})\frac{\frac{c-j+1}{c-j}}{(1-\beta_i\frac{c-j+1}{c})(\frac{c-j+1}{c-j}-\beta_i\frac{c-j+1}{c})}\prod_{i'\in[d-1]\setminus\{i\}} \frac{\beta_{i'}}{\beta_{i'}-\beta_i}\nonumber \\ &+g(P_{j-1})\frac{\frac{c-j+1}{c-j}}{\frac{c-j+1}{c-j}-1}\prod_{i\in[d-1]}\frac{\beta_i\frac{c-j+1}{c}}{\beta_i\frac{c-j+1}{c}-1}\nonumber\\&+g(P_{j})\frac{1}{1-\frac{c-j+1}{c-j}}\prod_{i\in[d-1]}\frac{\beta_i\frac{c-j+1}{c}}{\beta_i\frac{c-j+1}{c}-\frac{c-j+1}{c-j}}.
\end{align}
Recall the definition of $Q_j$, equation (\ref{eq:ex12}) is equivalently  
\begin{align}
 Q_j= \sum_{i\in[d-1]}g(\tilde{X}_{(i,j)})\frac{\frac{c-j+1}{c-j}}{(1-\beta_i\frac{c-j+1}{c})(\frac{c-j+1}{c-j}-\beta_i\frac{c-j+1}{c})}\prod_{i'\in[d-1]\setminus\{i\}} \frac{\beta_{i'}}{\beta_{i'}-\beta_i}.
\end{align}
Thus, $Q_j$ can be computed using a linear combination of $\{g(\tilde{X}_{(i,j)})\}_{i\in[d-1]}$. 

\section{Proof of Lemma \ref{lemma:mul}}\label{app:pl_lemma2}

%We prove Lemma \ref{lemma:mul} by induction. To enable an inductive structure, we consider a strictly stronger version, states as follows 

When $\textup{deg}\ g=1$, we need to prove that at least two workers are needed to compute $f$ with data-privacy.  The converse for this case is trivial, because we require that every single worker stores a random value that is independent of the input dataset, at least $1$ extra worker is needed to provide any information.   

Hence, we focus on the case where  $\textup{deg}\ g>1$. The rest of the proof relies on a converse bound we proved in \cite{yu2018lagrange} for a polynomial evaluation problem, stated as follows.

\begin{lemma}[Yu et al.]\label{lemma:lcc}
Consider a dataset with inputs $X_1,...,X_K$, and a multilinear function $g$ with degree $d\geq 1$. Any data-private linear encoding scheme that computes $g(X_1),...,g(X_K)$ requires at least $Kd+1$ workers.
\end{lemma}

Based on Lemma \ref{lemma:lcc}, we essentially need to prove that computing any gradient-type function with a multilinear $g$ with degree $d$ requires at least one more worker compared to evaluating multilinear functions with degree $d-1$. We prove this fact by constructing a degree $d-1$ multilinear function $g'$ and a corresponding scheme that evaluates $g'$ using $N-1$ workers, given any multilinear function $g$ with degree $d$ and any scheme for gradient-type computation which uses $N$ workers.  

Specifically, consider any fixed multilinear function $g:\bV_1\times...\times\bV_d\to \bU$ with $d$ inputs denoted by $X_{i,1},...,X_{i,d}$, we let $g'$ be a function that maps $\bV_1\times...\times\bV_{d-1}$ to a space of linear functions, such that $g'(X_{i,1},...,X_{i,d-1})(X_{i,d})=g(X_{i,1},...,X_{i,d})$ for any values of $X_{i,1},...,X_{i,d}$. One can verify that $g'$ is a multilinear map with degree $d-1$. 

%Then consider any fixed valid data-private encoding scheme for a gradient-type computation task specified by $g$ that uses $N$ workers, we aim to construct an encoding scheme for evaluating $g'$ using $N-1$ workers. For bervity, we refer to these schemes as $g$-scheme and  $g'$-scheme. Specifically, %fixing a subset of $N-1$ workers in the $g$-scheme,
%we let each worker $i$ in $g'$-scheme encodes the input variables as if encoding only the first $d-1$ entries  using the encoding function of worker $i$ in $g$-scheme.

Then consider any fixed valid data-private encoding scheme for a gradient-type computation task specified by $g$ that uses $N$ workers, we  construct an encoding scheme for evaluating $g'$ using $N-1$ workers, such that each worker $i$ encodes the input variables as if encoding only the first $d-1$ entries using the encoding function of worker $i$ in the gradient-type computation scheme.
For brevity, we refer to these two schemes as $g$-scheme and  $g'$-scheme. 

Due to the linearity of encoding functions, one can show that $g'$-scheme is data-private if $g$-scheme is data-private. Hence, it remains to prove the validity of $g'$-scheme. We first prove that for any fixed $X_{1,d},...,X_{K,d}$, $g'$-scheme can compute $\sum\limits_{i=1}^{K}g'(X_{i,1},...,X_{i,d-1})(X_{i,d})$. 

This decodability can be achieved in $2$ steps. First, due to the data-privacy requirement of $g$-scheme, given any fixed $X_{1,d},...,X_{K,d}$, one can find values of the padded random variables for the $d$th entry, such that the coded  $d$th entry stored by worker $N$ equals $0$. We denote the corresponding coded entry stored by any other worker $i$ by $\tilde{X}_{i,d}$. In this case, worker $N$ would return $0$ if $g$-scheme is used, due to the multilinearity of $g$. Second, given the results from $N-1$ workers, the master can evaluate them at points $\tilde{X}_{1,d},...,\tilde{X}_{N-1,d}$, respectively, which essentially recovers the computing results of workers $1,...,N-1$ in $g$-scheme. Knowing that the absent worker $N$ would return $0$, the needed gradient function  $\sum\limits_{i=1}^{K}g'(X_{i,1},...,X_{i,d-1})(X_{i,d})$ can thus be directly computed using the decoding function of $g$-scheme.

Now to recover each $g'(X_{i,1},...,X_{i,d-1})$ individually, it is equivalent to recover $g'(X_{i,1},...,X_{i,d-1})(X_{i,d})$ individually for any $X_{i,d}\in\bV_d$. This can be done by simply zeroing all the rest $X_{i',d}$'s, which completes the proof of validity.

Now that $g'$-scheme is valid and data-private, according to Lemma \ref{lemma:lcc}, it uses at least $K(d-1)+1$ workers. Hence $g$-scheme uses at least $K(d-1)+2$ workers. As the above proof holds true for any possible $g$-scheme, we have proved the converse bound in Lemma \ref{lemma:mul}.

\end{document}